\documentclass[12pt]{iopart}

\usepackage{iopams}
\usepackage{cite}
\usepackage{graphicx}
 \usepackage{mathrsfs}

\begin{document}
\title{Relativistic Klein-Gordon charge effects by information-theoretic measures.}

\author{D Manzano$^{1,2}$, R J Y\'a\~nez$^{1,3}$ and J S Dehesa$^{1,2}$}

\address{$^1$Instituto Carlos I de F\'isica Te\'orica y Computacional, Universidad de Granada,
18071-Granada, Spain}
\address{$^2$Departamento de F\'isica At\'omica, Molecular y Nuclear, Universidad de Granada, 18071-Granada, Spain}
\address{$^3$Departamento de Matem\'atica Aplicada, Universidad de Granada, 18071-Granada, Spain}
\eads{\mailto{manzano@ugr.es},\mailto{ryanez@ugr.es},\mailto{dehesa@ugr.es}}

\date{\today}

\begin{abstract}
The charge spreading of ground and excited states of Klein-Gordon particles moving in a Coulomb potential 
is quantitatively analyzed by means
 of the ordinary moments and the Heisenberg measure as well as by use of the most relevant information-theoretic
 measures of global (Shannon entropic power) and local (Fisher's information) types. The dependence of these
 complementary quantities on the nuclear charge $Z$ and the quantum numbers characterizing the physical states 
is carefully discussed. The comparison of the relativistic Klein-Gordon and non-relativistic Schr\"odinger values 
is made. The non-relativistic limits at large principal quantum number $n$ and for small values of $Z$ are also reached.
\end{abstract}

\maketitle

\section{Introduction}
The interplay of quantum mechanics, relativity theory and information theory is a most important topic in the present-day
theoretical physics \cite{AVE,PER,PLA,SEN,BOR,CUB,SAN}. Special relativity provokes both important restrictions on the transfer of
information between distant systems \cite{PER} and severe changes on the integral structure of physical systems \cite{GRE}.
This is mainly because the relativistic effects produce a spatial redistribution of the single-particle density $\rho(\vec{r})$ of 
the corresponding quantum-mechanical states, which substantially alter the spectroscopic and macroscopic properties of the 
systems.

The quantitative study of the relativistic modification of the spatial extent of the charge density of atomic and molecular
systems by information-theoretic means is a widely open field \cite{SEN}. The only works published up to now have calculated
 the ground-state relativistic effects on hydrogenic \cite{SEN} and many-electron neutral atoms \cite{BOR,SAN} in different 
settings by use of the renowned standard desviation (or Heisenberg measure) as well as various information-theoretic measures.

In this paper we quantify the relativistic effects of the ground and excited states of
 the spinless single-particle charge spreading by the comparison of the
 Klein-Gordon and Schr\"odinger values for three qualitatively different measures: the Heisenberg measure
  $\sigma\left[ \rho \right]$, the Shannon entropic power $N\left[ \rho \right]$ \cite{NIE} and 
the Fisher information $I\left[\rho \right]$ \cite{FRI,COV}. While the Heisenberg quantity gives the spreading with respect 
to the centroid of the charge distribution, the Shannon and Fisher measures do not refer to any specific point.

The Shannon entropic power $N\left[\rho \right]$, which is essentially given by the exponential of the Shannon 
entropy $S\left[ \rho \right]=-\left< \textrm{log}\;\rho(\vec{r})\right>$, measures the total extent to which the distribution 
is in fact concentrated \cite{SHA,COV}. This quantity has various relevant features. First, it avoids the dimensionality
troubles of $S\left[ \rho \right]$, highligting its physical meaning. Second, it exists when $\sigma$ does not. Third, it is
finite whenever $\sigma$ is. Thus, as a measure of uncertainty the use of the Shannon entropic power allows a wider 
quantitative range of applicability than the Heisenberg measure \cite{HAL}. Contrary to the Shannon
and Heisenberg  measures, which are insensitive to electronic oscillations, (translationally invariant) the Fisher information \cite{FRI} has a locality property because it is a gradient functional of the density, so that it measures the pointwise concentration of the electronic cloud and quantifies its gradient content, providing a quantitative estimation of the oscillatory character of the density. Moreover, the Fisher information measures the bias to particular points of the space, i.e. it gives a measure to the local disorder.

To calculate the measures of the charge spreading in a relativistic quantum-mechanical system we have to tackle the problem 
of the very concept of quantum probability consistent with Lorentz covariance. The general formulation and interpretation of 
this problem is still a currently discussed issue \cite{KIM2}. In this work we avoid this problem following the relativistic 
quantum mechanics \cite{GRE} by restricting ourselves to study the stationary states of a spinless relativistic particle with a 
negative electric charge in a spherically symmetric Coulomb potential $V(r)=-\frac{Z\; e^2}{r}$, which are the solutions of the 
relativistic scalar wave equation, usually called the Klein-Gordon equation \cite{SCH},

\begin{equation}
\left[ \epsilon-V(r) \right] \psi(\vec r)=( - \hbar^2 c^2 \nabla^2+m_0^2c^4 ) \psi(\vec r),
\end{equation}
appropriately normalized to the particle charge. The symbols $m_0$ and $\epsilon$ denote the mass and the relativistic energy 
eigenvalue, respectively. We will work in spherical coordinates, taking the ansatz $\psi(r,\theta,\phi)=r^{-1}u(r)Y_{lm}(\theta,\phi)$, 
where $Y_{lm}(\theta,\phi)$ denotes the spherical harmonics of order $(l,m)$. Then, to highlight the resemblance with the 
non-relativistic Schr\"odinger equation, we let

\begin{equation}
\beta \equiv \frac{2}{\hbar c} (m_0^2c^4-\epsilon^2)^\frac{1}{2}=\frac{2m_0c^2}{\hbar c}
\sqrt{1-\left( \frac{\epsilon}{m_0c^2} \right)^2},
\end{equation}

\begin{equation}
\lambda \equiv \frac{2 \epsilon Z e^2}{\hbar^2 c^2 \beta},
\end{equation}
and substitute the radial variable $r$ by the dimensionless variable $s$ through the transformation

\begin{equation}
r \to s:\qquad s=\beta r.
\end{equation}
So, the radial Klein-Gordon equation satisfied by $u(s)$ can be written in the form
\begin{equation}\label{eq:radial}
\frac{d^2 u(s)}{ds^2}-\left[ \frac{l'(l'+1)}{s^2}-\frac{\lambda}{s}+\frac{1}{4} \right] u(s)=0,
\end{equation}
where we have used the notation
\begin{equation}
l'=\sqrt{\left( l+\frac{1}{2}  \right)^2-\gamma^2}-\frac{1}{2},\qquad with \quad \gamma \equiv Z \alpha,
\end{equation}
being $\alpha=\frac{e^2}{\hbar c}$ the fine structure constant. The physical solutions corresponding to the bound 
states (whose energy eigenvalues fulfill $|\epsilon |<m_0c^2$) require that the radial eigenfunctions $u_{n l}(r)$ vanish 
both at the origin and at infinity \cite{NIE}, so that they have the form
\begin{equation}\label{eq:red}
u_{n l}(s)=\mathscr{N} s^{(l'+1)}e^{-\;\frac{s}{2}}\widetilde{L}_{n-l-1}^{2l'+1}(s),
\end{equation}
where $\widetilde{L}_{k}^{(\alpha)}(s)$ denotes the orthonormal Laguerre polynomials of degree $k$ and parameter
 $\alpha$. The energy eigenvalues $\epsilon \equiv \epsilon_{ln}(Z)$ of the stationary bound states with wavefunctions 
$\Psi_{nlm}(\vec r,t)=\psi_{nlm}(\vec r)exp(-\frac{i}{\hbar}\epsilon t)$ are known to have the form \cite{NIE}
\begin{equation}
\epsilon=\frac{m_0c^2}{\sqrt{1+\left( \frac{\gamma}{n-l+l'}\right)^2}}
\end{equation}
The constant $\mathscr{N}$ is determined not by the normalization of the wavefunction to unity as in the non-relativistic 
case, but by the charge conservation carried out by $\int_{\mathbb{R}^3}\rho(\vec{r})d^3r=e$ to preserve the Lorentz 
invariance \cite{GRE}, where the charge density of the negatively charged particle (e.g., a $\pi^-$-meson; $q=-e$) is given by
\begin{equation}\label{eq:densidad}
\rho_{n l m}(\vec{r})=\frac{e}{m_0 c^2}\left[ \epsilon-V(r)\right]  |\psi_{n l m}(\vec{r}) |^2.
\end{equation}
Then, the charge normalization imposes the following restriction on the radial eigenfunctions
\begin{eqnarray}\label{eq:norm}
1&=\int_0^{\infty} \frac{\epsilon -V(r)}{m_0c^2}u_{\epsilon l}^2(r)dr\nonumber\\
&=\frac{1}{m_0c^2} \int_0^{\infty}\left(\frac{\epsilon}{\beta}+
\frac{\gamma \hbar c}{\rho}  \right) u_{\epsilon l}^2(s)ds. 
\end{eqnarray}
The substitution of the expression (\ref{eq:red}) for $u_{\epsilon l}(s)$ into Eq. (\ref{eq:norm}) provides the following 
normalization constant
\begin{eqnarray}
\mathscr{N}^2&=m_0c^2\left[\frac{2\epsilon}{\beta}(n+l'-l)+\gamma\hbar c \right]^{-1}\nonumber\\
&=\frac{m_0c^2\gamma}{\hbar c} \frac{1}{(n+l'-l)^2+\gamma^2},
\end{eqnarray}
where we have used for the second equality the relation
\begin{equation}
\frac{\epsilon}{\beta}=\frac{\hbar c}{2}\frac{n+l'-l}{\gamma}
\end{equation}

\begin{figure}[!h]
\begin{center} 
\includegraphics[scale=0.5]{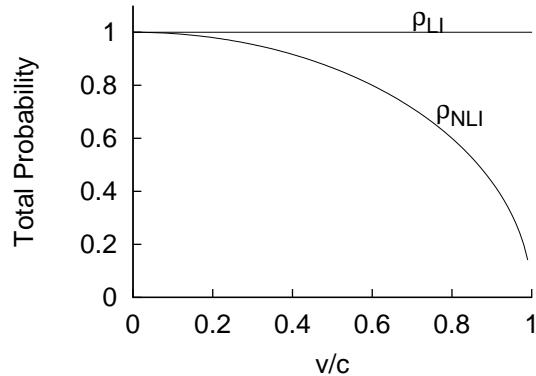}
\end{center}
\caption{Normalization of the charge density for the Lorentz invariant (LI) and the non-Lorentz invariant (NLI) charge densities
for different velocities of the observer.}
\label{fig:norma}
\end{figure}

Let us emphasize that the resulting Lorentz-invariant charge density $\rho_{LI}(\vec{r})$ given by Eq. 
(\ref{eq:densidad}) is always (i.e. for any observer's velocity $v$) appropriately normalized while the
 density $\rho_{NLI}(\vec{r})=   \left|\psi_{n l m}(\vec{r}) \right|^2$ (used in \cite{C06}) 
is not. This is numerically illustrated in Figure \ref{fig:norma}  for a pionic atom with nuclear charge $Z=68$ in the infinite 
nuclear mass approximation ($\pi^-$ -meson mass=273.132054 a.u.).

\begin{figure}[h]
\begin{center}
\includegraphics[angle=270,scale=.3]{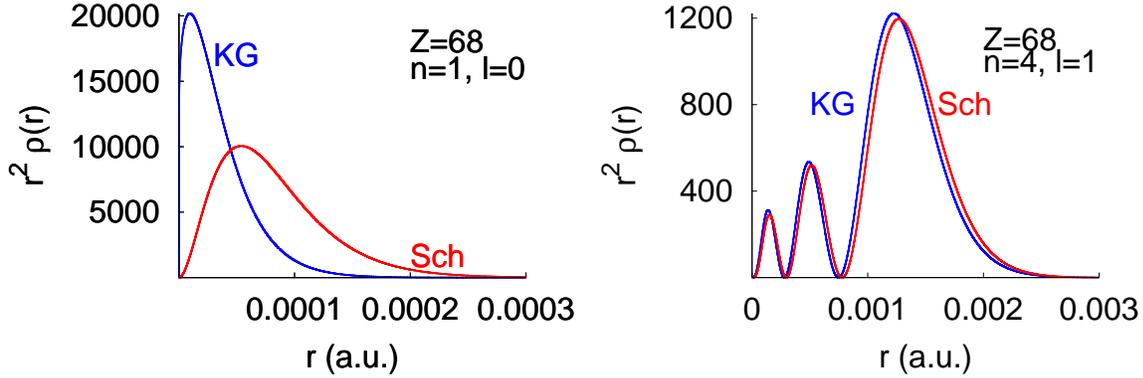}
\caption{\small{Comparison of the charge Klein-Gordon and Schr\"odinger radial density for the states $1S$ (left) and 
$4P$ (right) of the pionic system with $Z=68$. Atomic units ($\hbar=m_e=e=1$) are used.}}
\label{fig:density}
\end{center}
\end{figure}

For completeness we have plotted in Figure  \ref{fig:density} the radial density of the charge distribution for two diferent states 
$(n=1,\;l=0)$ and $(n=4,\;l=1)$ of a pionic system with nuclear charge $Z=68$ in the infinite nuclear mass 
approximation, respectively. Moreover, we have also made in these figures 
a comparision with the corresponding Schr\"odinger density functions \cite{SHE}. We observe that the relativistic 
effects other than spin (i) tend to compress the charge towards the origin, and (ii) they are most apparent for states $S$. 

In this paper we quantify this relativistic charge compression by three different means. First, in Section II, we compute 
the ordinary moments or radial expectation values  $\left<r^k\right>$ for general $(n,\;l,\;m)$ states, making emphasis in the 
Heisenberg measure for circular $(l=n-1)$ and $S$-states $(l=0)$. Then, in Section III, we study numerically the most 
relevant charge information-theoretic measures of the system; namely the Shannon entropy and the Fisher information.

\section{Radial expectation values and Heisenberg's measure.}
\label{sec2}

The charge distribution of the Klein-Gordon particles in a Coulomb potential can be completely characterized by means 
of the ordinary radial expectation values $\left< r^k \right>$, $k\in \mathbb{N}$, given by

\begin{eqnarray}
\left<r^k\right>&:= \int_{\mathbb{R}^3} r^k \rho_{n l m}(\vec r) d^3 r\nonumber\\
&=\frac{1}{m_0 c^2} \int_0 ^\infty \left(  \epsilon+\frac{Ze^2}{r}  \right) r^k u_{n l}^2(r) dr\nonumber \\
&=\frac{1}{m_0c^2} \frac{1}{\beta^{k}}\int_0^\infty \left(\frac{\epsilon}{\beta}+
\frac{\gamma \hbar c}{s} \right) s^k u_{n l}^2(s) ds \nonumber\\
&=\frac{\mathscr{N}^2}{m_0c^2} \frac{1}{\beta^{k}}\left[\frac{\epsilon}{\beta} \mathscr{J}_{nl}(k)+
\gamma \hbar c \mathscr{J}_{nl}(k-1)\right],
\end{eqnarray}

where we have used Eqs. (\ref{eq:red}) and (\ref{eq:densidad}), and the symbol $\mathscr{J}_{nl}(k)$ denotes
 the integral \cite{NIE}
\newpage
\begin{eqnarray}
\mathscr{J}_{nl}(k)&:=\int_0^\infty x^{2l'+k+2}e^{-x}\left[\widetilde L_{n-l-1}^{(2l'+1)}(x)\right]^2dx\nonumber\\
&=\frac{(n-l-1)!}{\Gamma(n-l+2l'+1)}\nonumber\\
&\times \sum_{j=n-l-k-2}^{n-l-1}\; \left(\begin{array}{c} k+1\\ n-l-j-1 \end{array}\right) ^2\frac{\Gamma(2l'+k+j+3)}{j!}
\end{eqnarray}
For the lowest values of $k$ we have

\begin{eqnarray}
\mathscr{J}_{nl}(0)&=2(n+l'-l)\nonumber\\
\mathscr{J}_{nl}(1)&=2\left[3(n-l)^2+l'(6n+2l'-6l-1)\right] \nonumber\\
\mathscr{J}_{nl}(2)&=4(n+l'-l)\nonumber\\
&\quad \times \left[1+5(n-l)^2+l'(10n+2l'-10l-3)\right].\nonumber
\end{eqnarray}

Then, besides the normalization $\left<r^0\right>=1$, we have the following value

\begin{equation}\label{eq:r}
\left<r\right>=\frac{\mathscr{N}^2}{m_0c^2}\frac{1}{\beta}\left[\frac{\epsilon}{\beta} \mathscr{J}_{nl}(1)+
\gamma \hbar c \mathscr{J}_{nl}(0) \right]
\end{equation}
for the centroid of the charge density, and

\begin{equation}\label{eq:r2}
\left<r^2\right>=\frac{\mathscr{N}^2}{m_0c^2}\frac{1}{\beta^2}\left[\frac{\epsilon}{\beta} \mathscr{J}_{nl}(2)+
\gamma \hbar c \mathscr{J}_{nl}(1) \right]
\end{equation}
for the second-order moment, so that the Heisenberg measure $\sigma_{nl}$ which quantifies the charge 
spreading around the centroid is given by

\begin{eqnarray}
\sigma^2_{nl}\equiv \sigma\left[\rho_{nlm}\right]&=\left<r^2\right>-\left<r\right>^2\nonumber\\
& =\frac{\mathscr{N}^2}{m_0 c^2}\frac{1}{\beta^2} \left\{\frac{\epsilon}{\beta} \mathscr{J}_{nl}(2)+ 
\gamma \hbar c  \mathscr{J}_{nl}(1)-\right.\nonumber\\
&\quad -\left. \frac{\mathscr{N}^2}{m_0 c^2}\left[ \frac{\epsilon}{\beta} \mathscr{J}_{nl}(1)+
\gamma \hbar c \mathscr{J}_{nl}(0)  \right]^2  \right\}.
\end{eqnarray}
To gain insight into these general expressions we are going to discuss two particular classes of 
quantum-mechanical states, the circular (i.e., $l=n-1$) states and the  ns-states (i.e., $l=0$) states.

For circular states we have that

\begin{displaymath}
\begin{array}{l}
l'=\sqrt{\left( n-\frac{1}{2} \right)^2-\gamma^2}-\frac{1}{2}\\
\epsilon=\frac{m_0c^2}{\sqrt{1+\left( \frac{\gamma}{l'+1} \right)^2}};
\frac{\epsilon}{\beta}=\frac{\hbar c}{2 \gamma}(l'+1),\\
\end{array}
\end{displaymath}
so that

\begin{displaymath}
\frac{\mathscr{N}^2}{m_0 c^2}=\frac{\gamma}{\hbar c}\frac{1}{\left( l'+1  \right)^2+\gamma^2},
\end{displaymath}
and the integrals

\begin{displaymath}
\begin{array}{l}
\mathscr{J}_{nl}(0)=2l'+2\\
\mathscr{J}_{nl}(1)=(2l'+2)(2l'+3)\\
\mathscr{J}_{nl}(2)=(2l'+2)(2l'+3)(2l'+4)\\
\end{array}
\end{displaymath}

Then, the centroid of the charge distribution is, according to Eq. (\ref{eq:r}),

\begin{equation}
\left<r\right>=\frac{\hbar c}{4 m_0 c^2}\frac{1}{\gamma\sqrt{1+\left( \frac{\gamma}{l'+1} \right)^2}}
\left[ (2l'+2)(2l'+3)+4 \gamma^2   \right]
\end{equation}
and the second-order moment, according to equation (\ref{eq:r2}), becomes

\begin{equation}
\left<r^2\right>=\left( \frac{\hbar c}{m_0c^2} \right)^2 \frac{1}{2 \gamma^2} (l'+1)(2l'+3)\left[ (l'+1)(l'+2)+
\gamma^2  \right],
\end{equation} 
so that the Heisenberg measure for circular states $\sigma_n^2=\sigma_{n,n-1}^2$ has the following value

\begin{eqnarray}
\sigma_n^2&=\left( \frac{\hbar c}{m_0c^2} \right)^2 \left(\frac{l'+1}{4 \gamma^2}\right) \nonumber\\
&\quad \times \frac{(l'+1)(2l'+3)\left[ (l'+1)^2+2\gamma^2 \right]+2\gamma^4}{(l'+1)^2+\gamma^2}.
\end{eqnarray}

\begin{figure}[!ht]
\begin{center}
\includegraphics[angle=270,scale=0.3]{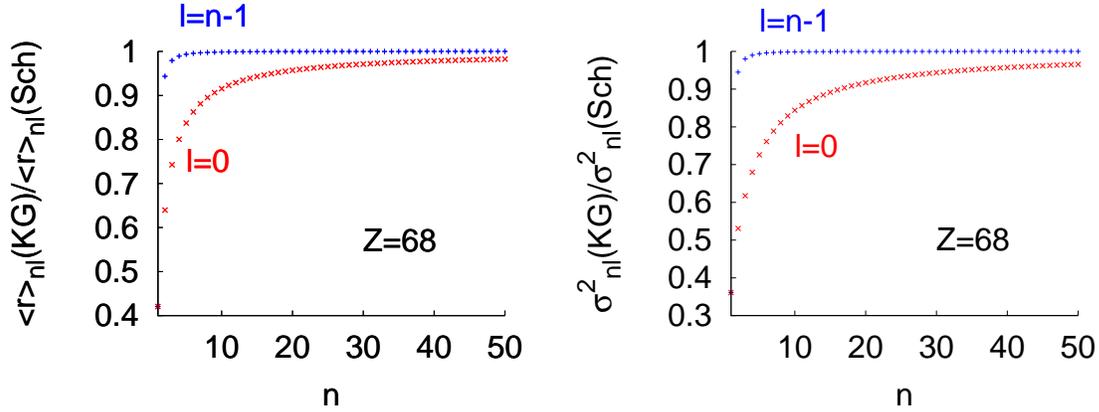}
\caption{\small{ Comparison of the Klein-Gordon and Schr\"odinger values for the centroid (Left) and the variance 
(Right) of the pionic $1S$ and the circular states as a function of the quantum number $n$ with $Z=68$.}}
\label{fig:radiovarianceN}
\end{center}
\end{figure}

\begin{figure}[!h]
\begin{center}
\includegraphics[angle=270,scale=0.3]{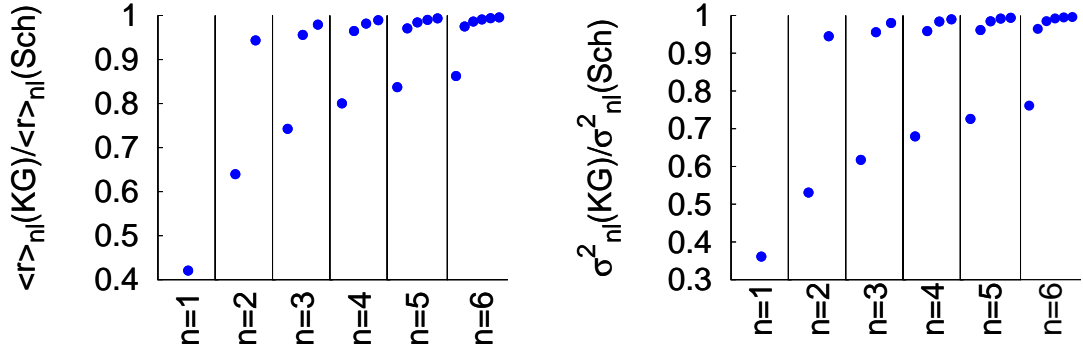}
\caption{\small{Comparison of the Klein-Gordon and  Schr\"odinger values for the centroid (Left) and the variance 
(Right), as a function of the quantum number $l$ varying from $0$ to $n-1$, for different values of $n$.}}
\label{fig:radiovarianceL}
\end{center}
\end{figure}

These expressions for circular states and the corresponding ones for ns-states are discussed and compared 
with the Schr\"odinger values as a function of the principal quantum number n for the pionic system with 
nuclear charge $Z=68$ in Figure \ref{fig:radiovarianceN}. We observe that both centroid and variance ratios increase very rapidly 
with n, being the rate of this behaviour much faster for circular than for $S$-states. This indicates that the 
charge compression provoked by relativity in a given system (i.e. for fixed $Z$) (i) decreases when $n$ $(l)$ 
is increasing for fixed $l$ $(n)$. This can also be noticed in Figure \ref{fig:radiovarianceL}, where the two previous ratios have been 
plotted as a function of $l$ for different values of $n$.

\begin{figure}[!ht]
\begin{center}
\includegraphics[angle=270,scale=0.3]{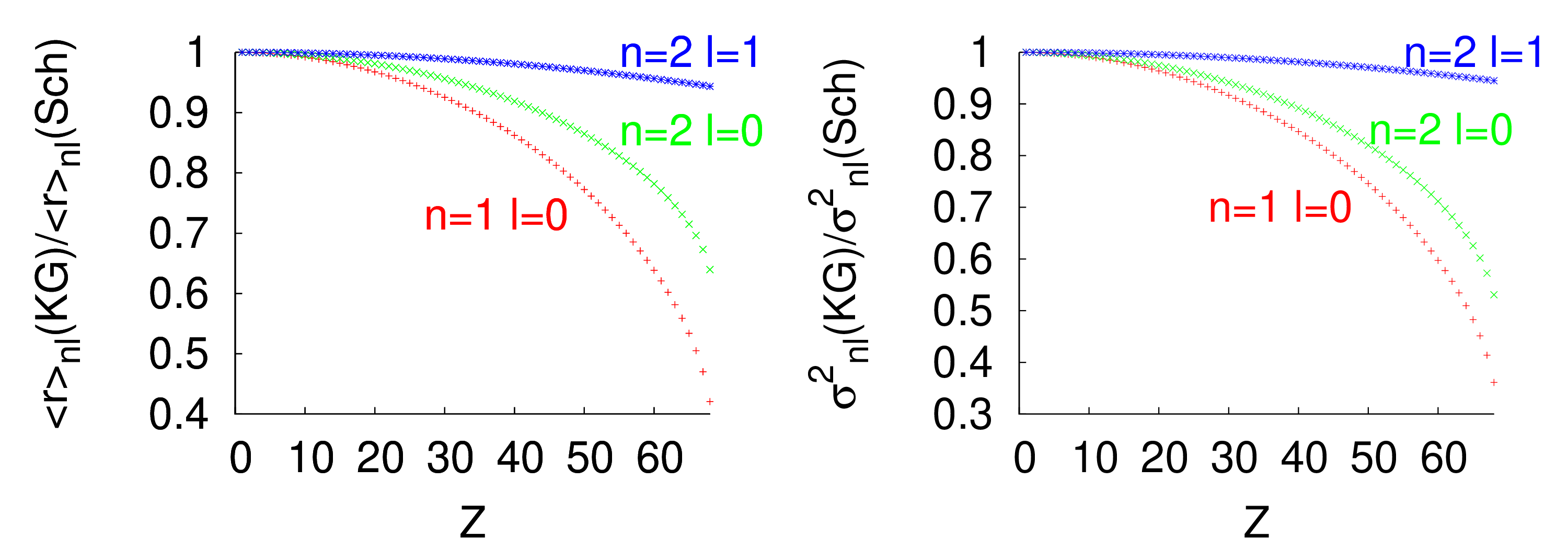}
\caption{\small{Comparison of the Klein-Gordon and  Schr\"odinger values for the centroid (Left) and the variance 
(Right), as a function of the nuclear charge $Z$ for the pionic states $1S$, $2S$ and $2P$.}}
\label{fig:radiovarianceZ}
\end{center}
\end{figure}

  Then, we have plotted these two ratios in terms of the nuclear charge $Z$ of the system in Figure \ref{fig:radiovarianceZ} for the 
states $1S$, $2S$ and $2P$. We find that both the centroid and the variance ratios monotonically decrease as 
the nuclear charge $Z$ increases. Moreover, the decreasing rate is much faster for the states $1S$, than for 
the states $2S$ and $2P$. These two observations illustrate that the relativistic charge compression effect  is 
bigger in heavier systems for a given ($nl$)-state. Moreover, we see here again that for a given system it 
increases both when $n$ decreases for fixed $l$ and when $l$ decreases for fixed $n$. The quantum number 
$m$ doesn't affect both ratios because the radial part of the density is not a function of it.

Finally, let highlight that in all figures  the Klein-Gordon values tend towards the Schr\"odinger values in the non-relativistic 
limit of large $n$ or small $Z$.

\section{Shannon and Fisher information measures}
\label{sec3}

Here we study numerically the relativistic effects  on the charge spreading of pionic systems of hydrogenic type by means 
of the following information-theoretic measures of the associated charge distribution $\rho_{nlm}(\vec{r})$ given by 
Eq (\ref{eq:densidad}): The Shannon entropy power and the Fisher information.

The Shannon entropic power of a negatively-charged Klein-Gordon particle characterized by the charge density 
$\rho_{nlm}(\vec{r})$ is defined by \cite{COV}
\begin{equation}
N_{nlm}\equiv N\left[\rho_{nlm}\right]=\frac{1}{2\pi e}\textrm{exp}\left(\frac{2}{3}S_{nlm}\right),
\end{equation}
where $S_{nlm}$ is the Shannon entropy of $\rho_{nlm}(\vec{r})$ given by the expectation value of 
$-\textrm{log}\left( \rho_{nlm}(\vec{r}) \right)$, i. e.
\begin{equation}\label{eq:shannonentropy}
S_{nlm}\equiv S\left[\rho_{nlm}\right]=-\int_{\mathbb{R}^3} \rho_{nlm}(\vec{r})\; \textrm{log}\; \rho_{nlm}(\vec{r}) d^3 r,
\end{equation}
which quantifies the total extent of the charge spreading of the system. Taking into account the above-mentioned ansatz for $\psi(\vec{r})$  and Eqs. (\ref{eq:red}), (\ref{eq:densidad}) and (\ref{eq:shannonentropy}), this expresion can be separated out into radial and 
angular parts 
\begin{displaymath}
S_{nlm}=S\left[R_{nl}  \right]+S\left[Y_{lm}\right],
\end{displaymath}
as it is explained in full detaill in \cite{SHE2}, being $R_{nl}$ and $Y_{lm}$ the radial and angular parts of the density. We should keep in mind that the angular part is the 
same for both Klein-Gordon and Schr\"odinger cases. 

The  Fisher information is defined by \cite{FRI}
\begin{equation}
I_{nlm}\equiv I\left[\rho_{nlm}\right]=\int_{\mathbb{R}^3}\frac{|\nabla \rho(\vec{r}) |^2}{\rho(\vec{r})} d^3 r.
\end{equation}
Remark that we are not using here the parameter dependent Fisher information originally introduced (and so much used) by  statisticians \cite{fish}, but its translationally invariant form that does not depend on any parameter; see ref.  \cite{FRI,HM09} for further details.
It is worthy to point out that the Fisher information is a measure of the gradient content of the charge distribution: so, when 
$\rho(\vec{r})$ has a discontinuity at a certain point, the local slope value drastically changes and the Fisher information 
strongly varies. This indicates that it is a local quantity in contrast to the Heisenberg measure $\sigma_{nl}^2$ and the 
Shannon entropy $S_{\rho}$ (and its associated power), which have a global character because they are powerlike and 
logarithmic functionals of the density, respectively.

 Unlike the moment-based quantities discussed in the previous section, these complementary measures do not depend on a 
special point, either the origin as the ordinary moments or the centroid as the Heisenberg measure. These quantities, first used 
by statisticians and electrical engineers and later by quantum physicists, have been shown to be measures of disorder or 
smoothness of the density $\rho_{nlm}(\vec{r})$  \cite{FRI, COV}. Let us highlight that the Fisher information does not only 
measure the charge spreading of the system in a complementary and qualitatively different manner as the Heisenberg and Shannon 
measures but also it quantifies their oscillatory character, indicating the local charge concentration all over the space \cite{FRI}.

\begin{figure}[!h]
\begin{center}
\includegraphics[angle=270,scale=0.3]{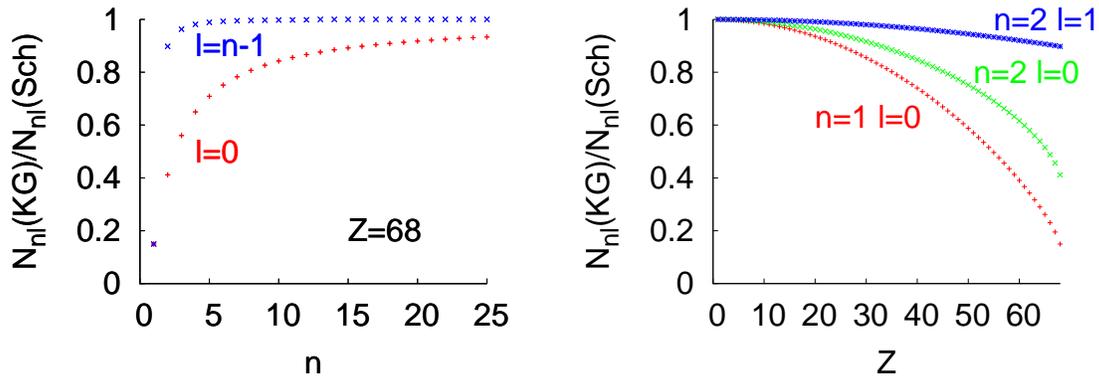}
\caption{\small{Comparison of the Klein-Gordon and  Schr\"odinger values for the Shannon entropic power as a function of the 
principal quantum number $n$ (Left) and the nuclear charge $Z$ (Right).}}
\label{fig:shannon}
\end{center}
\end{figure}

The relativistic (Klein-Gordon) and non-relativistic (Schr\"odinger) values of the Shannon entropic power are numerically discussed 
and compared in Figure \ref{fig:shannon}  for the pionic system. Therein, on the left, we plot the ratio 
$N_{nl}(\textrm{KG})/N_{nl}(\textrm{Sch})$  between these 
two values as a function of the principal quantum number $n$ for the system with nuclear charge $Z=68$. We notice that the 
Shannon ratio systematically increases when $n$ is increasing, approaching to unity for large n, for both circular and $S$-states. 
Moreover, we find that this approach is much faster for circular states, what indicates once more that the relativistic effects are 
much more important for states S. In addition, on the right of Figure \ref{fig:shannon}, we show the dependence of the Shannon ratio with the 
nuclear charge $Z$ for the $1S$, $2S$ ant $2P$ states. We observe, here again, that the ratio is a decreasing function of $Z$ for 
any state, indicating that the relativistic effects are much more important for heavy systems. Moreover, for a given system 
(i.e. fixed $Z$) the relativistic effects increase when $n$ $(l)$ decreases for fixed $l$ $(n)$. The quantum number $m$ affects the 
absolute value of the Shannon entropic power but it doesn't affect the ratio.

\begin{figure}[!h]
\begin{center}
\includegraphics[angle=270,scale=0.3]{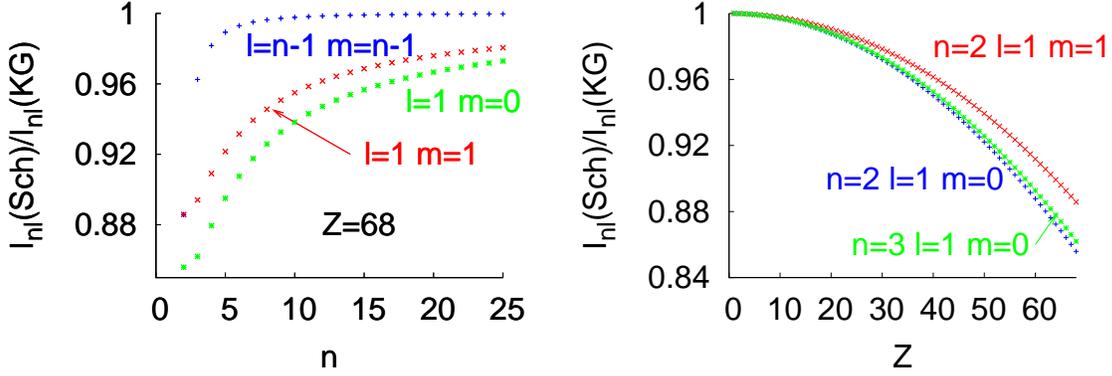}
\caption{\small{Comparison of the Klein-Gordon and  Schr\"odinger values for the Fisher information as a function of the principal 
quantum number $n$ (Left) and the nuclear charge $Z$ (Right).}}
\label{fig:fisher}
\end{center}
\end{figure}

\begin{figure}[!h]
\begin{center}
\includegraphics[angle=270,scale=0.3]{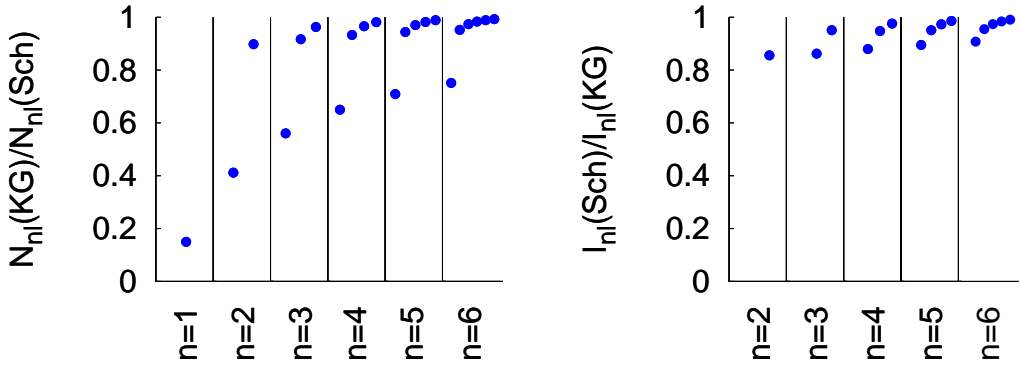}
\caption{\small{Comparison of the Klein-Gordon and  Schr\"odinger values for the Shannon entropic power as a function of $l$ variying 
from $0$ to $n-1$ (Left) and the Fisher information as a function of $l$ variying from $1$ to $n-1$ (Right) for different values of $n$.}}
\label{fig:shannonfisherratio}
\end{center}
\end{figure}

Figure \ref{fig:fisher} shows the dependence of the ratio of the non-relativistic and relativistic values of the Fisher information for various states with 
$l\neq 0$ on their quantum numbers $(n,l,m)$ for the pionic system with $Z=68$ (left graph) and on the nuclear charge $Z$ (right graph). 
The Fisher information for $S$-states is not defined because the involved integral diverges. First we should remark that here, contrary 
to the previous quantities considered in this work, the Schr\"odinger values are always less than the Klein-Gordon ones; this is strongly 
related to the local character of the Fisher information, indicating that the localized internodal charge concentration is always larger in 
the relativistic case. Second, we observe that for fixed $l$ the Fisher ratio 
$I_{nl}(\textrm{Sch})/I_{nl}(\textrm{KG})$ monotonically increases when $n$ is 
getting bigger, approaching to unity at a rate which grows as $l$ is increasing. Third, we find that the Fisher ratio decreases for all states in 
a systematic way as the nuclear charge increases. Moreover, for a given $Z$ value this ratio increases as either the quantum numbers 
$n$ and/or $l$ increase.

For completeness, the behavior of the Shannon and Fisher ratios in terms of the orbital quantum number $l$ for a fixed $n$ is more explicitly shown of the left and right graphs, respectively, of Figure \ref{fig:shannonfisherratio}.

\begin{figure}[!h]
\begin{center}
\includegraphics[angle=270,scale=0.3]{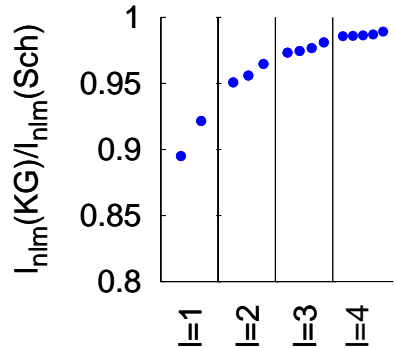}
\caption{\small{Comparison of the Klein-Gordon and  Schr\"odinger values for the Fisher information as a function of $m$ varying 
from $0$ to $l$, for different values of $l$.}}
\label{fig:fisherM}
\end{center}
\end{figure}

Finally, in Figure \ref{fig:fisherM},  the dependence of the Fisher ratio on the magnetic quantum number $m$ is studied. Notice that the ratio is bigger when $|m|$ is increasing, 
indicating that the lower $|m|$ is, the more concentrated is the charge density of the state and the more important are the relativistic effects.

\section{Conclusions}
\label{sec4}

The relativistic charge compression of spinless Coulomb particles has been quantitatively investigated by means of the Heisenberg, 
Shannon and Fisher spreading measures. These three complementary quantities show that the relativity effects are larger (i. e. the 
charge compresses more towards the origin) for the lower energetic states and when the Coulomb strength (i. e. the nuclear charge 
$Z$) increases. Moreover, a detailed analysis of these quantities on the quantum numbers $(n,l,m)$ characterising the physical 
states of a given system (i. e. for a fixed $Z$) indicate that the relativistic effects increase when $n$ $(l)$ decreases for fixed $l$ $(n)$.
 Furthermore, the study of the Fisher information shows that the relativistic effects also increase when the magnetic quantum 
number $|m|$ is increasing for fixed $(n,l)$.

\section{Acknowlegments}
We are very grateful to Junta de Andalucia for the grants FQM-2445 and FQM-1735, and the Ministerio de Ciencia e 
Innovaci\'on for the grant FIS2008-02380/FIS. We belong to the research group FQM-207. Daniel Manzano acknowledges the 
fellowship BES-2006-13234.


\newpage
\begin{thebibliography}{99}
\bibitem{AVE} J. Avery, Information Theory and Evolution (World Sci. Publ., N.Y.,2003).
\bibitem{PER} A. Peres and D.R. Terno, Rev. Mod. Phys. \textbf{76} 93  (2004).
\bibitem{PLA} A.R. Plastino and C. Zander, in A Century of Relativity Physics: ERE2005, AIP Conference Proceedings, \textbf{841}, 570 (2006).
\bibitem{SEN} J. Katriel, K.D. Sen,  J. Comput.  Appl. Math. (2009), 
doi:10.1016/j.cam.2008.04.039.
\bibitem{BOR} A. Borgoo, F. de Prost, P. Geerlings and K.D. Sen, Chem. Phys. Lett. \textbf{444} (2007) 186.
\bibitem{CUB} D. Cubero J. Casado-Pascual, J. Dunkel, P. Talkner and P. H\"anggi. 
Phys. Rev. Lett. \textbf{99} (2007) 170601; D. Cubero and J. Dunkel, Europhys. Lett. \textbf{87} (2009) 30005.
\bibitem{SAN} J. Sa\~nudo and R. L\'opez-Ruiz, Phys. Lett. A \textbf{373} (2009) 2549.
\bibitem{GRE} W. Greiner, {\it Relativistic Quantum Mechanics: Wave Equations}, 3rd edition (Springer, Berlin, 2000).
\bibitem{NIE} M.M. Nieto, Am. J. Phys. \textbf{47} 1067 (1979). Attention to the typographical error in Eq. (5.12).
\bibitem{FRI} B.R. Frieden, {\it Science from Fisher information}. (Cambridge University Press, Cambridge, 2004).
\bibitem{fish} R.A. Fisher, Proc. Camb. Phil. Soc. \textbf{22} (1925) 700.
\bibitem{COV} T.M. Cover and J. A. Thomas, {\it Elements of Information Theory} (Wiley, N. Y., 1991).
\bibitem{SHA} C. E. Shanon, Bell. Sys. Tech. \textbf{27} (1948) 379 and 623 [reprinted in Claude Elwood
Shannon: Collected Papers, edited by N. Sloane and A. Wyner (IEEE, New York, 1993)]; A.J. Stam, 
Information and Control \textbf{2} (1959) 101.
\bibitem{HAL} M.J.W. Hall, Phys. Rev. A \textbf{59} (1999) 2602.
\bibitem{KIM2} Y.S. Kim and M. E. Noz, Found. Probab. Phys.-4 \textbf{889} 152 (2006). See also arXiv:quant-ph/030115 (2003).
\bibitem{SCH} E. Schr\"odinger, Ann. Phys. 81 (1926) 109; O. Klein, Z. Phys. \textbf{37} 895 (1926); V. Fock, Z. 
Phys. \textbf{38} 242  (1926); ibid \textbf{39} 226 (1926) ; W. Gordon, Z. Phys. \textbf{40} 117 (1926).
\bibitem{C06} C.Y. Chen and S.H. Dong, Phys. Scr. \textbf{73} 511 (2006).
\bibitem{SHE} J.S. Dehesa, S. L\'opez-Rosa, A. Mart\'inez-Finkelstein and R. J. Y\'a\~nez, Int. J. Quantum Chem. (2009). Accepted.
\bibitem{SHE2} J.S. Dehesa, S. L\'opez-Rosa, B. Olmos, R. J. Y\'a\~nez, J. Comput. Appl. Math. \textbf{179} 
185-194 (2005).
\bibitem{HM09} I.P. Hamilton, R.A. Mosna, J. Comput. Appl. Math.  doi:10.1016/j.cam.2009.02.087  (2009).
\end{thebibliography}
\end{document}